\documentclass[11pt]{article}
\usepackage{epsfig,sint}

\newcommand{\C}{\rm \kern.25em\vrule height1.4ex
depth-.12ex width.06em\kern-.31em C}

\newcommand{\rme}{{\rm e}}
\newcommand{\rmd}{{\rm d}}
\newcommand{\rmO}{{\rm O}}
\newcommand{\GeV}{{\rm GeV}}

\newcommand{\be}{\begin{equation}}   
\newcommand{\ba}{\begin{eqnarray}}
\newcommand{\ea}{\end{eqnarray}}

\newcommand{\wt}{{\widetilde{w}}}
\newcommand{\Mt}{{\widetilde{M}}}
\newcommand{\Wt}{{\widetilde{W}}}

\newcounter{subequation}[equation]
\makeatletter

\expandafter\let\expandafter
\reset@font\csname reset@font\endcsname

\def\subeqnarray{\arraycolsep1pt
    \def\@eqnnum\stepcounter##1{\stepcounter{subequation}%
        {\reset@font\rm(\theequation\alph{subequation})}}
\jot5mm     \eqnarray}

\makeatother
\newcommand{\msbar}{{\rm \overline{MS\kern-0.14em}\kern0.14em}}

\begin{document}

\begin{titlepage}

\begin{flushright}
   MPP-2004-43\\
   May 2004
\end{flushright}

\vskip 0.20 true cm

%\vskip 0.20 true cm
\begin{center}
{\Large\bf 
Structure functions of $2d$ integrable asymptotically free models}
\end{center}
\vskip 1 true cm
\centerline{\large Janos Balog}
\vskip1ex
\centerline{Research Institute for Particle and Nuclear Physics}
\centerline{1525 Budapest 114, Pf. 49, Hungary}
\vskip 1 true cm
\centerline{\large Peter Weisz}
\vskip1ex
\centerline{Max-Planck-Institut f\"ur Physik}
\centerline{F\"ohringer Ring 6, D-80805 M\"unchen, Germany}
\vskip 1 true cm
\centerline{\bf Abstract}
\vskip 1.0ex
We investigate structure functions in the 2--dimensional 
(asymptotically free) non--linear O$(n)\,\,\sigma$--models 
using the non--perturbative S--matrix bootstrap program.  
In particular the {\it exact small (Bjorken) $x$ behavior is exhibited};
the structure is rather universal and is probably the 
same in a wide class of (integrable) asymptotically free models. 
Structurally similar universal formulae may also hold 
for the small $x$ behavior of QCD in 4--dimensions. 
Structure functions in the special case of the $n=3$ model are accurately 
computed over the whole $x$ range for $-q^2/M^2<10^5$, and some moments
are compared with results from renormalized perturbation theory.
Some remarks concerning the structure functions in the $1/n$  
approximation are also made.

\vfill
\eject

\end{titlepage}

%%%%%%%%%%%%%%%%%%%%%%%%%%%%%%%%%%%%%%%%%%%%%%%%%%%%%

\section{Introduction}

Structure functions describing scattering of electrons and neutrinos
off nucleon targets are well measured and give us insight into
the structure of the nucleons \cite{Buras,Altarelli,Roberts,FoRo}.
At high $-q^2$ and intermediate
Bjorken $x$ the parton model and DGLAP equations \cite{DGL,AP} give
a good description. At smaller $x$ the DLGAP equations are considered to
break down and BFKL--type \cite{BFKL} equations take over, here the
structure function $F_2\sim x^{-\nu(-q^2)}$ with $\nu(-q^2)>0$. 
A value of $\nu(0)>0$ would however (seem to) violate the Froissart bound.
Recently saturation models, such as the so--called color glass model
\cite{colourglass1,colourglass2} predict the true asymptotic behavior 
to be $F_2\sim\ln^p x\,,$ with $p=1$ or 2. 

The QCD literature on small $x$ physics is vast and rather involved
\cite{Smallxcollab,Mueller}. One certain aspect is that a
description of the asymptotically small $x$ region 
requires some crucial non--perturbative input.
The most systematic non--perturbative methods for QCD, using the lattice 
regularization, are able to give non--perturbative information
on the moments of the structure functions via the OPE
\cite{QCDSF,Negele}, however they are not applicable to yield 
information on the asymptotically small $x$ behavior.

In this paper we study structure functions in asymptotically free 
integrable models in two dimensions. Despite the fact that there
are no transverse directions, the structure functions have a rather rich 
and non--trivial behavior with many features reminiscent of the
structure functions in QCD. Here we will concentrate on results obtained
for O$(n)$ sigma models. 
In particular we have derived the {\it exact small 
$x$} behavior in these models. The result has a rather universal
model--independent structure, being independent of $n$ and holding
for a large class of operator probes. One is tempted to speculate a
similar qualitative structure to hold in QCD.

\section{Sigma model 2--point functions}

The O$(n)$ $\sigma$--model in $2d$ formally described by the Lagrangian
\be
{\cal L}=\frac{1}{2g_0^2}\sum_{a=1}^n
\partial_\mu S^a\partial^\mu S^a
\,,\,\,\,\,\,\,\,\,\sum_{a=1}^n S^aS^a=1\,, 
\end{equation}
is perturbatively asymptotically free for $n\ge3$.  
A special property is that these models have
an infinite number of local \cite{Polyakov}
and non--local \cite{Luscher} classical conservation 
laws which survive quantization. 
At the quantum level they imply absence of particle production. 
Assuming the spectrum to consist of one stable 
O$(n)$--vector multiplet of mass $M$,
the S--matrix has been proposed long ago by the Zamolodchikovs \cite{ZZ}.
Form factors of local operators can be computed 
using general principles \cite{Smirnov}. 
The S--matrix bootstrap program for the construction of correlation 
functions involves summing the contributions over all intermediate states
\cite{Karowski}. The possible equivalence of this construction
to the continuum limit of the lattice regularized theory has been
investigated in ref.~\cite{SigmaI}. In papers 
of one of the present authors (J.B) and M.~Niedermaier \cite{JanosMax}
2--point functions of
various operators were computed, including those of the O($n$) current
$J_\mu^{cd}$ and spin--field operators $\Phi^a$. 
We give their definitions here because they will be
needed later to state our result on the small $x$ behavior:
\ba
&&\langle0|T\,J^{cd}_\mu(x)J^{ef}_\nu(y)|0\rangle=
\nonumber\\
&&-i\left(\delta^{ce}\delta^{df}-\delta^{cf}\delta^{de}
\right)\int\frac{\rmd^2 q}{(2\pi)^2}
\rme^{-iq(x-y)}\left(q_\mu q_\nu-q^2\eta_{\mu\nu}\right)I_1(-q^2-i\epsilon)\,,
\label{current2ptfn}
\ea
valid up to contact terms and
\be
\langle0|T\Phi^a(x)\Phi^b(y)|0\rangle=-i\delta^{ab}\Lambda_n^2
\int\frac{\rmd^2 q}{(2\pi)^2}  
\rme^{-iq(x-y)}I_0(-q^2-i\epsilon)\,,
\label{spin2ptfn}
\end{equation}
where the normalization factor $\Lambda_n$ is chosen (for later convenience):
\be
\Lambda_3=\frac{2}{\sqrt\pi}\,,\,\,\,\,\,\Lambda_n=1\,\,\,\,\,n>3\,.
\end{equation} 
To complete the definitions we must supplement the operator normalizations.
The currents have a normalization fixed by requiring their spatial 
integrals to yield the correct charges (cf Eq.~(\ref{currentnorm})),
and our field normalization is fixed by requiring
\be
\langle 0|\Phi^a(0)|b,\theta\rangle=\delta^{ab}
\end{equation}
for one particle states with momentum $p_1=M\sinh\theta$, 
with state normalization
\be
\langle a,\theta'|b,\theta\rangle
=4\pi\delta^{ab}\delta(\theta'-\theta)\,.
\end{equation}
The studies of these 2--point functions (in the case $n=3$) 
\cite{JanosMax} presently constitute the best 
evidence for the existence of a non--perturbative construction 
of a model with asymptotic freedom. 

\section{Sigma model structure functions}

The $\sigma$--model analogue of the central object in deep inelastic
scattering is
\be
W^{ab;cdef}_{\mu\nu}(p,q)={1\over 4\pi}\int\rmd^2x e^{iqx}
\langle a,p|\Bigl[J_{\mu}^{cd}(x),J_{\nu}^{ef}(0)\Bigr]|b,p \rangle\,.
\label{wmunu}
\end{equation}
We define the usual DIS kinematic variables
\be
\nu=pq/M\,,\,\,\,\,\,\,\,
x=-q^2/(2M\nu)\,,\,\,\,\,\,\,\,W^2=(p+q)^2\,.
\end{equation}
In the relevant kinematic domain $q^2<0, W\ge M$ i.e.
$0\le x\le1$ we have 
\be
W^{ab;cdef}_{\mu\nu}(p,q)=
\pi\sum_r \langle a,p|J_{\mu}^{cd}(0)|r\rangle
\langle r|J_{\nu}^{ef}(0)| b,p \rangle\delta^{(2)}(p+q-P_r)\,,
\label{wmunu1}
\end{equation}
where the sum is over the complete set of $r$--particle (``in" or ``out") 
states. Using Lorentz and O$(n)$ invariance we have
\be
W^{ab;cdef}_{\mu\nu}(p,q)=
\left(\eta_{\mu\nu}-\frac{q_{\mu}q_{\nu}}{q^2}\right)
\sum_{l=0}^2 w_l(q^2,x)R_l^{ab;cdef}\,,
\end{equation}
with projectors $R_l$ corresponding to the 3 invariant $t$--channel
``isospins" given by
\ba
R_0^{ab;cdef}&=&\frac{1}{n}\delta^{ab}
(\delta^{ce}\delta^{df}-\delta^{cf}\delta^{de})\,,
\\
R_1^{ab;cdef}&=&\frac12(\delta^{ac}\delta^{be}-\delta^{bc}\delta^{ae})
\delta^{df}
-(c\leftrightarrow d)-(e\leftrightarrow f)+(c\leftrightarrow
d\,,e\leftrightarrow f),
\\
R_2^{ab;cdef}&=&-\frac12\Bigl\{(\delta^{ac}\delta^{be}+\delta^{bc}\delta^{ae})
\delta^{df}
-(c\leftrightarrow d)-(e\leftrightarrow f)+(c\leftrightarrow
d\,,e\leftrightarrow f)\Bigr\}
\nonumber\\
&&+\frac{2}{n}\delta^{ab}
(\delta^{ce}\delta^{df}-\delta^{cf}\delta^{de})\,.
\ea
Note in 2 dimensions there is only one structure function for each isospin
channel, since there is only one (conserved symmetric) tensor
involving two momenta; one has e.g. ( for $p^2=M^2$)
\be
\Bigl(p_{\mu}-{M\nu q_{\mu}\over q^2}\Bigr)
\Bigl(p_{\nu}-{M\nu q_{\nu}\over q^2}\Bigr)
=\frac{M^2(\nu^2-q^2)}{(q^2)^2}
\left( q_{\mu}q_{\nu}-\eta_{\mu\nu}q^2\right)\,.
\end{equation}

Although in QCD structure functions for operators not 
associated with physical sources have so far not been studied, 
we also introduce, for reasons which will soon become apparent, 
the field structure functions through
\ba                     
\Sigma^{ab;cd}(p,q)&=&-\pi q^2\sum_r 
\langle a,p|\Phi^c(0)|r\rangle \langle r|\Phi^d(0)| b,p \rangle
\delta^{(2)}(p+q-P_r) 
\\
&=&\Lambda_n^2\sum_{l=0}^2 \wt_l(q^2,x)P_l^{ab;cd}\,,
\ea
with $t$--channel projectors $P_l$ for the vector representation given by
\ba
P_0^{ab;cd}&=&\frac{1}{n}\delta^{ab}\delta^{cd}\,,
\label{proj0} 
\\
P_1^{ab;cd}&=&\frac{1}{2}\Bigl(\delta^{ac}\delta^{bd}    
-\delta^{bc}\delta^{ad}\Bigr)\,,
\label{proj1}
\\
P_2^{ab;cd}&=&\frac{1}{2}\Bigl(\delta^{ac}\delta^{bd}
+\delta^{bc}\delta^{ad}\Bigr)-\frac{1}{n}\delta^{ab}\delta^{cd}\,.
\label{proj2}
\ea
The current operators connect only states with an even number
of particles to the vacuum and the field operators only states 
with and odd number: 
\be
w_l(q^2,x)=\sum_{r\,\,{\rm odd}}w_l^{(r)}(q^2,x)\,,\,\,\,\,\,\,\,\,
\wt_l(q^2,x)=\sum_{r\,\,{\rm even}}w_l^{(r)}(q^2,x)\,.
\end{equation}

The wonderful feature of integrable models is that one knows
many explicit properties concerning the form factors appearing in
the expressions. These are encapsulated in the Smirnov axioms \cite{Smirnov},
and using these one can derive the previously announced exact   
asymptotic small $x$ behavior of the structure functions:
\ba
w_l(q^2,x)\sim\frac{1}{x\ln^2x}\frac{2\pi T_l}{(n-2)^2}A_1(-q^2)\,,
\label{smallx1}
\\
\wt_l(q^2,x)\sim\frac{1}{x\ln^2x}\frac{2\pi V_l}{(n-2)^2}A_0(-q^2)\,,
\label{smallx2}
\ea
where the functions $A_s$ are the ``Adler functions" defined
from the vacuum 2--point functions 
(\ref{current2ptfn}),(\ref{spin2ptfn}) by
\be
A_s(z)=-z^2\frac{\partial}{\partial z}I_s(z)\,,\,\,\,\,\,s=0,1\,.
\end{equation}
The constants appearing in (\ref{smallx1}),(\ref{smallx2}) are 
characteristic of the O$(n)$
representations (vector $V$ and anti--symmetric tensor $T$) of the 
corresponding operators: 
\ba
V_0=2(n-1)\,,\,\,\,\,\,V_1=V_2=n-2\,,
\\
T_0=4(n-2)\,,\,\,\,\,\,T_1=n-2\,,\,\,\,\,T_2=4-n\,.
\ea 
The explicit proof of these relations will be presented in a 
forthcoming more detailed publication \cite{JanosPeter2}. 
In this letter we would like to concentrate on general features. 
Firstly we note that the structure of the asymptotic small $x$ 
behavior is independent of the operator, independent of $n$, 
and independent of the channel. Further since the results were 
obtained from rather general principles, 
we think that they are valid for a large class of 
integrable asymptotically free models. Indeed we have checked that
the same behavior holds in the leading orders of the $1/n$
expansion in the Gross--Neveu model \cite{GrNe}. 

In the derivation of (\ref{smallx1}),(\ref{smallx2}) it appears that
the $1/\ln^2x$ behavior is related to the high energy behavior
of the scattering amplitudes; this is similar to the association 
of the proposed $\ln^p x$ in QCD with the asymptotic behavior of 
total cross sections (related to the forward scattering amplitude 
through the optical theorem). 

The question is what can one learn for QCD; is the small $x$ behavior
there also given by a structural formula factorizing a part characteristic
to the target and a part described by the vacuum 2--point function?
Unfortunately so far we have not succeeded to derive such a general
result. As a first guess we have looked at the ratio
of the structure function $F_2$ to the (electro--magnetic) 
Adler function $D$ in QCD, 
with HERA data at some of the lowest $x$ values published so far
\cite{H1}, and the Adler function taken from ref.~\cite{EJKV}. 
The result is presented in Fig.~1;
there is no sign that the ratio is becoming independent of $-q^2$
as $x$ decreases.
However from such comparisons one should be cautious to draw 
conclusions concerning the asymptotic behavior,
because at these values of $x$ the situation 
is qualitatively similar (only the slope is different)
for the ratio $xw_0(q^2,x)/A_1(-q^2)$ in the O(3) $\sigma$--model, 
which is also plotted in the same figure.
In this model one would have to go to much smaller values of $x$. 
The question remains for QCD, at which value of 
$x$ (for a given $-q^2$ range) does the true asymptotic behavior set in?

\begin{figure}[t]
\begin{center} 
%\vspace*{-1mm}
\hspace{1.0cm}   
\epsfig{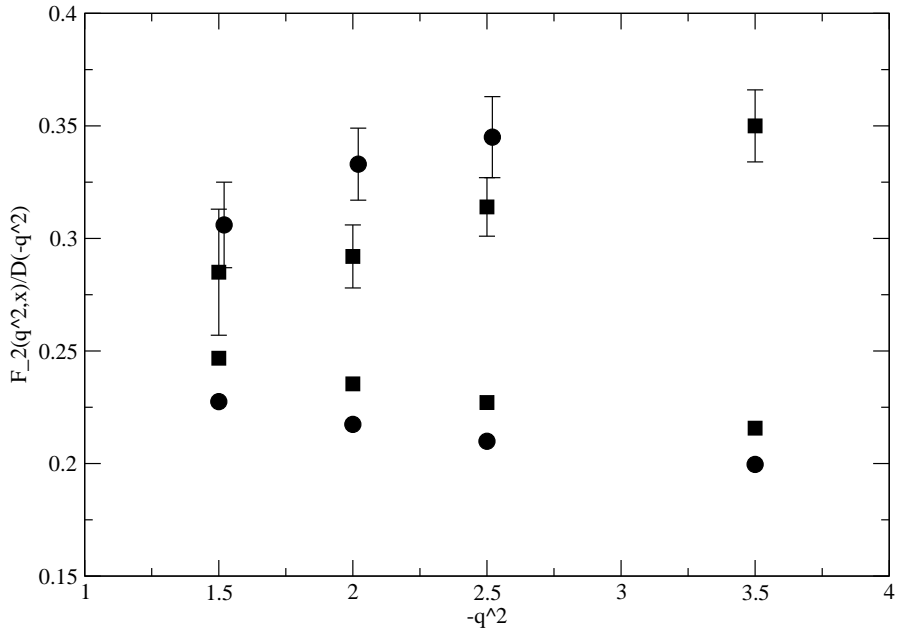}
\caption{\footnotesize
$F_2(q^2,x)/D(-q^2)$ for $x=5\cdot 10^{-5}$ (circles with error 
bars) and $x=8\cdot 10^{-5}$ (squares with error bars), 
for various values of $-q^2$ (in $\GeV^2$). 
Symbols without error bars are values of $xw_0(q^2,x)/A_1(-q^2)$
in the O(3) $\sigma$--model at the same values of $x$
(here the abscissa denotes values of $-q^2/M^2$).  
}
%\vspace*{-10mm}
\end{center}  
\label{FigAdler}
\end{figure}

The QCD structure functions in the range of small $x$ between $10^{-5}$ 
and $10^{-2}$ are fitted quite well with a ``Lipatov form"
$A(-q^2)x^{-\nu(-q^2)}$. As an exercise we have made least--square 
fits of $xw_0(q^2,x)$ (for O(3)) with such a form and
observed that in the regime  $10^{-5}\le x\le 10^{-3}$ such fits 
with $\nu(-q^2)=-0.192,\,-0.167,\,-0.151$ for $-q^2/M^2=1,10,100$
respectively, describe the data better than simplest fits
of the form $C(-q^2)/\ln^2x$ incorporating the known 
asymptotic small $x$ behavior.

\subsection{$1/n$ expansion}

The data obtained for O(3) used above and in subsequent sections  
require a considerable amount of computation. For this reason
we include here a short discussion of structure functions in
leading order of the $1/n$ expansion, where many qualitative results
are rather similar to $n=3$ but where relatively simple analytic 
formulae are available. 
The easiest case is the leading order of the spin structure functions
in the isospin 1,2 channels because they are given by the 
imaginary part of the propagator of the ``auxiliary field":
\ba
\wt_l(q^2,x)&=&\frac{1}{n}\theta(W-2M)4\pi\frac{-q^2M^2}{(-q^2+M^2)^2}
\frac{\sinh\theta}{\theta^2+\pi^2}+\rmO(1/n^2)\,,
\nonumber\\
&&{\rm for}\,\,\,W^2=4M^2\cosh^2\frac{\theta}{2}\,,\,\,\,l=1,2\,.
\ea
Apart from suggesting that the limits $n\to\infty$ and $x\to0$ commute, 
this simple function already illustrates many 
rather general features of the structure functions in this model. 
Firstly that the onset of the limit $x\to0$ is not uniform in $-q^2$. 
In the Bj--limit $-q^2\to\infty \,,\,\,\,\,x\,\,{\rm fixed}$ we have
\be
\wt_l(q^2,x)\sim\frac{1}{n}\frac{2\pi(1-x)}{x\ln^2(-q^2/M^2)}
\,,\,\,\,\,\,l=1,2.
\end{equation}
Secondly the limit $x\to0$ for fixed $q^2$ is approached extremely
slowly e.g. for $-q^2/M^2=1\,,\,\,\wt_l/(\wt_l)_{\rm asympt}
=0.93$ for $x=10^{-5}$, while for 
$-q^2/M^2=100$ we have e.g. $\wt_l/(\wt_l)_{\rm asympt}=0.49,0.59$
for $x=10^{-5},10^{-7}$ respectively.

We remark that the leading order (in $1/n$) isospin 0 (field) structure 
function involves another (box) diagram which is also easily evaluated. 
One can show that the small $x$ behavior is as predicted by the 
general formula (\ref{smallx2}). 
The leading $1/n$ diagram contributing to the current structure
functions is just a 1--particle exchange and thus only contributes 
a term $\propto\delta(1-x)$. We have computed the next--to--leading
orders for the $l=1,2$ channels and again confirm the predicted behavior.

\subsection{The case $n=3$}

So far we have concentrated on the small $x$ region; in the following
we extend our description of the structure functions
to the whole range of $x$. We will do this for the case $n=3$ which 
is rather special for various reasons. For our studies
in the S--matrix bootstrap approach it is important that it is 
the model for which the $r$--particle form factors 
can most easily be obtained explicitly. Moreover the spin and
current 2--point functions exhibit in this case very similar features
and there are miraculous scaling relations \cite{JanosMax} which relate 
them 
\footnote{see also the OPE in sect.~4.2}. 
There are well defined recursive 
procedures for computing the form factors, the only 
practical limitation being that they become very involved.
So far the record we have achieved is the 7--particle form factor
of the spin field \cite{JanosPeter1}; already its algebraic expression in 
MAPLE involves many megabytes of storage. Fortunately for the structure 
functions we only require sums over bilinear factors of the form factors
which are computationally more manageable.

Just as for the 2--point functions \cite{JanosMax}
we find that for a fixed $-q^2$ only
states with a limited number of particles contribute significantly.
To appreciate this better we consider the sum of the field and current
structure functions, which is a rather peculiar thing to do 
in general, but which is in fact rather natural in the special case $n=3$.
Figs.~2 and 3 illustrate how the structure 
function $x(w_0+\wt_0)$ is built up from 
states with increasing particle number for the cases $-q^2/M^2=10^2$ and 
$-q^2/M^2=10^4$ respectively. We see that the higher states 
contribute very little and
that we obtain nearly exact values for the structure functions
for all values of $-q^2/M^2<10^5$ by including only
intermediate states with $\le 5$ particles for the current 
and $\le 6$ particles for the spin field.

\begin{figure}[t]
\begin{center} 
%\vspace*{-1mm}
\hspace{1.0cm}   
\epsfig{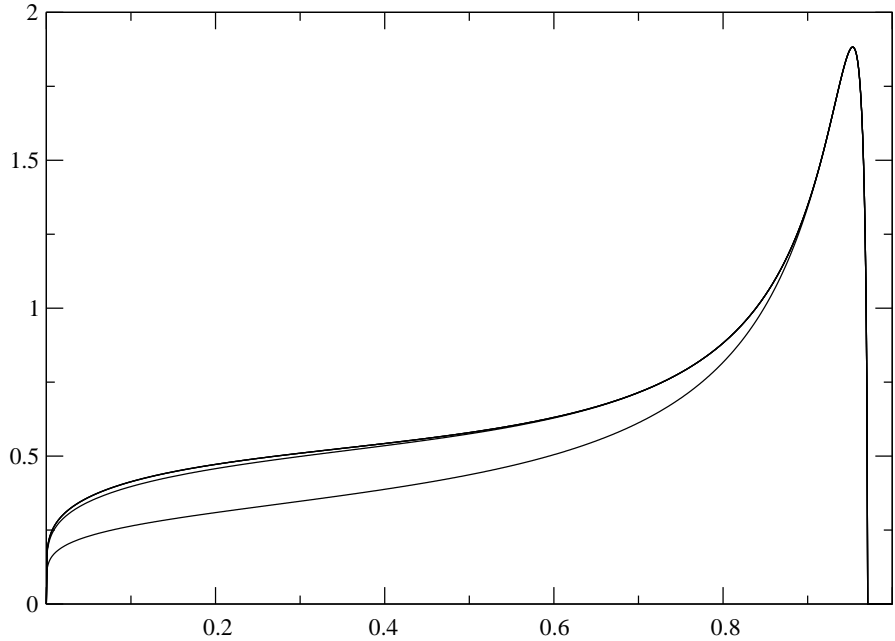}
\label{wwtilde100}
\caption{\footnotesize Approximations to
$x(w_0+\wt_0)$ as functions of $0<x<1$ for $-q^2/M^2=100$.
Curves correspond to sums up to and including $2,3,4,5,6$--particle 
intermediate states. The last 3 curves are indistinguishable on this 
scale. 
}
%\vspace*{-10mm}
\end{center}  
\end{figure}

\begin{figure}[t]
\begin{center} 
%\vspace*{-1mm}
\hspace{1.0cm}   
\epsfig{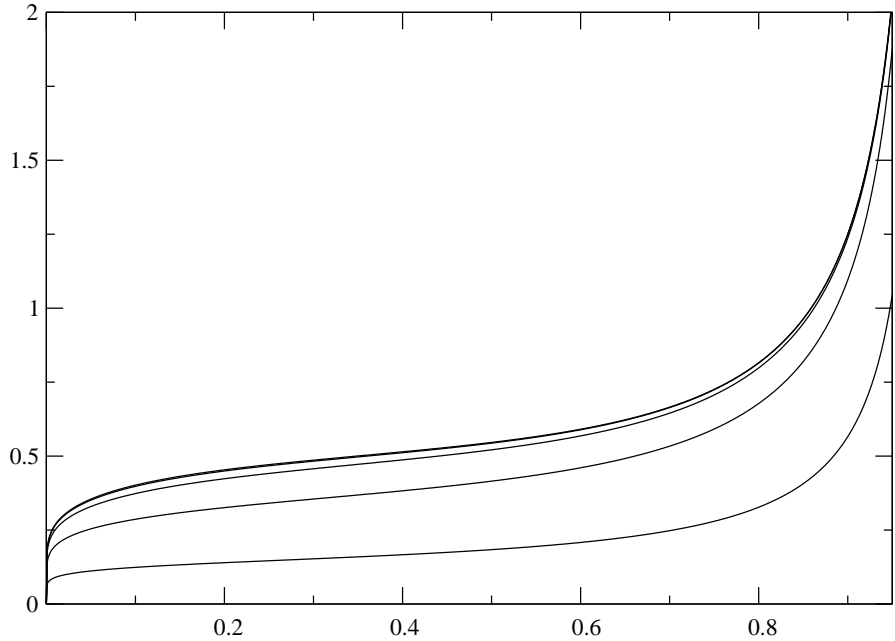}
\label{wwtilde10000}
\caption{\footnotesize Approximations to
$x(w_0+\wt_0)$ functions of $0<x\le0.95$ for $-q^2/M^2=10^4$.
Curves correspond to sums up to and including 2,3,4,5,6--particle states.
}
%\vspace*{-10mm}
\end{center}  
\end{figure}

\begin{figure}[t]
\begin{center}
%\vspace*{-1mm}
\hspace{1.0cm}
\epsfig{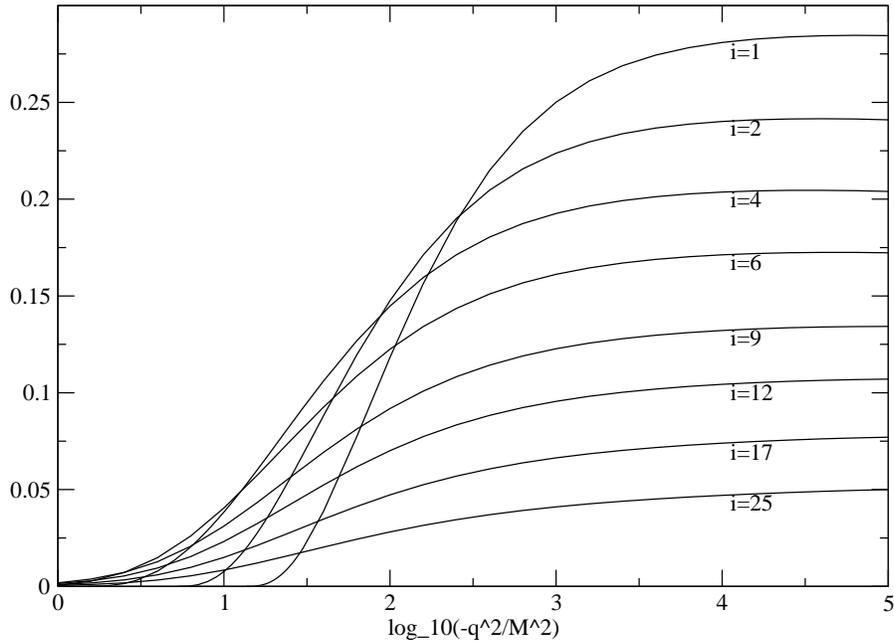}
\label{xw0HERA}
\caption{\footnotesize $xw_0(q^2,x)$ for various values of $x=10^{-i/5}$.
}                                                      
%\vspace*{-10mm}                           
\end{center}
\end{figure}
In Fig.~4 we plot $xw_0(q^2,x)$ as a function
of ${\rm log}_{10}(-q^2/M^2)$, for a selection of $x$--values
\footnote{For this model we prefer to show this rather 
than the typical HERA plot
where one adds $-\log_{10}(x)$ to separate the $x$--values,
because the latter would obscure the $-q^2$ variation which is rather
small compared to the variation of $-\log_{10}(x)$.}.
The function increases as $-q^2$ increases for all values of $x$ in this 
range.

\subsection{Threshold behavior}

Note that in Fig.~3 we have cut off the plot at 
$x=0.95$. This is because near $x=1$ the function develops a big bump 
with a peak $\sim 70$ which, if included in the same plot, would
obscure the features we wanted to show there. 
The behavior of the $\sigma$--model structure functions
near $x=1$ is indeed rather involved. For a fixed $-q^2$ 
the contribution to the structure function
from the $r$--particle state $w^{(r)}$ vanishes for $x$ greater
than some threshold value 
\be
x_r(-q^2)=\left[1-(r^2-1)M^2/q^2\right]^{-1}\,.
\end{equation}  
The big bump in $x(w_0+\wt_0)$ referred to above is at this value of 
$-q^2/M^2=10^4$ practically entirely due to the $2$--particle 
contribution. For this contribution: 
\ba
w^{(2)}_l&\sim& E_l(q^2)\sqrt{x_2(-q^2)-x}\,,\,\,\,x\to x_2\,,\,\,\,-q^2
\,\,\,\,{\rm fixed}\,,
\\
w^{(2)}_l&\sim& F_l(x)/\ln^2(-q^2/M^2)\,,\,\,\,-q^2\to\infty\,,\,\,\,x
\,\,\,\,{\rm fixed}\,,
\ea
where $E_l,F_l$ are some (known) functions. The bump arises because
$F_l$ is quite singular near threshold, $F_l\sim [(1-x)\ln^2(1-x)]^{-1}$. 
The analytic behavior as $x\to x_2$ sets in only extremely close
to threshold e.g for $-q^2/M^2=10^4$ 
the position of the peak of the bump is 
at $x=0.99954$ whereas the function vanishes at $x_2=0.99970$.
At $-q^2/M^2=10^4$ the 3--particle contribution also has a bump 
but it is less pronounced; (peak value $\sim 2.5$ at $x\sim0.9953$). 
We conjecture that the threshold behavior
of $w_0^{(r)}$ in the O(3) model is $(x_r-x)^{(r^2-3)/2}$.

One can also study the threshold behavior in the $n\to\infty$ limit.
Here we find that in the leading order $\wt_l\sim \sqrt{x_2(-q^2)-x}$
for $l=1,2$ but $\wt_0\sim 1/\sqrt{x_2(-q^2)-x}$. We caution 
however that the limits $n\to\infty$ and $x\to$ threshold 
may not commute.

The characteristic enhanced near--threshold features in this model 
probably have no counterpart in QCD. For QCD the behavior of the structure
functions in the  $x\sim1$ region is surely also complicated. 
But there the effects might be quite suppressed for large $-q^2$
since, if we model the behavior in this region by the contribution
of resonances, their electromagnetic form factors are thought 
to fall very quickly (as powers) in $-q^2$ similarly to that 
of the proton. 

\section{Partons, OPE and moments}

\subsection{Parton model}

In the O$(n)\,\,\sigma$--models there does not seem to be a simple parton
picture. This is even so for the case $n=3$ where the model is equivalent
to the $\C{\rm P}^1$ model. For although this model is formulated in 
terms of a complex doublet of fields which are analogous to  
quarks in that they are confined, it seems that they do not play a r\^{o}le 
more similar to partons than the elementary bare spin fields in the 
original formulation
\footnote{Perhaps the peculiar threshold behavior discussed in sect.~3.3 
is explained by the fact that (as opposed to QCD) with some probability
the O$(n)$ particle can consist of a single point--like
parton that carries the same quantum numbers.}. 
The question is related to that of understanding  
what are (if any) the ``ultra--particles" in the sense 
of Buchholz and Verch \cite{BuVe}, 
or to the associated question as to whether the 
$\sigma$--models have an underlying conformal field theory. 

Although an intuitive parton description with suggestive DGLAP equations
\be
q^2\frac{\partial}{\partial q^2}w_l(q^2,x)
=\int_x^1\frac{\rmd y}{y} p_l(x/y,q^2)w_l(q^2,y)\,,
\label{ateqtn}
\end{equation}
(where $p_l(z,q^2)$ would be the corresponding splitting functions) 
is still missing in these models, we still have the machinery of the 
operator product expansion (OPE) which we apply in the following.

\subsection{Moments}

A class of interesting quantities are the moments 
of the structure functions:
\ba
M_{l;N}(q^2)&=&\int_0^1\rmd x x^{N-1} w_l(q^2,x)
=\sum_{r={\rm odd}}M_{l;N}^{(r)}(q^2)\,,
\\
\Mt_{l;N}(q^2)&=&\int_0^1\rmd x x^{N-1} \wt_l(q^2,x)
=\sum_{r={\rm even}}M_{l;N}^{(r)}(q^2)\,,
\ea
where $M_{l;N}^{(r)}(q^2)$ denote the contributions from $r$--particle 
states
\be
M_{l;N}^{(r)}(q^2)=\int_0^1\rmd x x^{N-1} w_l^{(r)}(q^2,x)\,.
\end{equation}
As in QCD the moments satisfy renormalization group equations
from which one can determine their leading behavior as $-q^2\to\infty$.
These are derived by considering the OPE for the 
current--current and field--field products and explicitly treating
the terms involving the operators of highest (zero) twist 
\footnote{``Twist" in this model is the naive dimension minus the
``spin" ($=$number of uncontracted Lorentz indices).}.
The general analysis is rather involved because the classification
of lowest twist operators turns out to be more complicated
than in QCD because the elementary field is dimensionless
\footnote{cf. in QCD the quark field carries dimension 3/2}.
Our analysis extends that initiated 
e.g. in refs.\cite{Luscher}, \cite{CMP}. 
Here we just quote some results and defer the derivations
to ref.~\cite{JanosPeter2}. 

For the current ($N$ even) moments in the isospin 0 channel we have
\begin{equation}
M_{0;N}(q^2)=W_{0;N}\,\frac{n-2}{2(n-1)}
\,\left\{1+\frac{1}{n-2}\lambda(q^2)+\rmO\left(\lambda^2\right)\right\}\,,\,\,\,
N\ge2\,,
\label{curr0J}
\end{equation}
where $\lambda(q^2)$ is an effective running coupling function defined 
through
\be
\frac{1}{\lambda(q^2)}+\frac{1}{n-2}\ln\lambda(q^2)
=\ln \frac{\sqrt{-q^2}}{\Lambda_\msbar}\,,
\end{equation}
and the $W_{0;N}$ are
renormalization group invariant, non-perturbative constants,
corresponding to the matrix elements of spin $N$ operators. 
In the $N=2$ case this is the energy--momentum tensor 
operator $T_{\mu\nu}$ for which we know the constant explicitly
\be
\langle a,\theta|T_{++}(0)|b,\theta\rangle=
\frac14 W_{0;2}M^2\rme^{2\theta}\delta^{ab}\,,\,\,\,\,W_{0;2}=2\,, 
\end{equation}
where the index $+$ means ``$(0-1)/2$". In particular
the ``momentum sum rule" follows:
\be
M_{0;2}(-\infty)=\frac{n-2}{n-1}\,.
\label{m02infty}
\end{equation}
Note that all the isospin 0 moments tend to constants as 
$-q^2\to\infty$.
As a consequence these current structure functions in the O($n$) models 
obey Bjorken scaling, and Fig.~4 indicates that the
resulting limiting scaling functions are non--trivial.
This is a special property of these models and 
we conjecture that this is due to the existence of
an infinite set of local conserved quantities \cite{Polyakov}. 

In the isospin $l=1$ channel for odd moments $N\ge3$ 
we can only say that
\begin{equation}
M_{1;N}(q^2)=W_{1;N}\,\lambda(q^2)^{\frac{1}{n-2}}+\dots\,,\,\,N\ge 3\,,
\label{curr1J}
\end{equation}
but in the special case $N=1$ we have
\begin{equation}
M_{1;1}(q^2)=\frac{1}{2}\,
\,\left\{1-\frac{1}{n-2}\lambda(q^2)+\rmO\left(\lambda^2\right)\right\}\,,
\label{curr11}
\end{equation}
where the constant is known through the current normalization
\be
\langle 
a,\theta|J_+^{cd}(0)|b,\theta\rangle=-2iM\rme^{\theta}P_1^{ab;cd}\,.
\label{currentnorm}
\end{equation}
From this follows the analogy to the Adler sum rule in QCD:
\be
M_{1;1}(-\infty)=\frac12\,.
\label{m11infty}
\end{equation}

For the spin field isospin 0 moments we have 
\be
\Mt_{0;N}(q^2)
=\cases{\frac{W_{0;N}\pi^2nC_n}{(n-2)^2}\lambda(q^2)^{\frac{n-3}{n-2}}
\left\{1+\rmO(\lambda)\right\}\,,\,\,\,\,n\ge4\,;\cr
\frac{W_{0;N}}{4}\left\{1+\lambda(q^2)+\rmO(\lambda^2)\right\}
\,,\,\,\,\,\,\,\,\,\,\,\,\,\,\,\,\,\,n=3,\cr}
\label{spin0J}
\end{equation}
where the non--perturbative constants $W_{0;N}$ are the same as for the 
current, and where $C_n$ is the non--perturbative constant appearing 
in the short distance expansion 
\be
\langle0|\Phi^a(y)\Phi^b(0)|0\rangle\sim 
C_n\delta^{ab}\left(-\ln M|y|\right)^{\frac{n-1}{n-2}}\,,
\end{equation}
which is only known for the cases $n=3$ and $n=\infty$ ($C_\infty=1/(2\pi)$).
We see that only for the case $n=3$ do the moments of the 
field $l=0$ structure function have the same leading asymptotic 
behavior as those of the current.

For the isospin $l=1$ field (odd) moments we find to leading orders PT  
\be
\Mt_{1;N}(q^2)
=\cases{\Mt_{0;2}(q^2)\,,\,\,\,\,\,\,\,\,\,\,\,\,\,\,\,\,\,\,\,
\,\,\,\,N=1\,;\cr
\Wt_{1;N}\lambda^{\frac{2n-5}{n-2}}+\dots\,,\,\,N\ge3,\cr}
\label{spin1J}
\end{equation}
where there is in general no obvious relation between the 
$\Wt_{1;N}$ and the constants occurring in (\ref{curr1J}), except 
for $n=3$ where they are equal ($\Wt_{1;N}=W_{1;N}\,,\,\,n=3)$.

In Figs.~5 and 6 we plot the separate $r$--particle
contributions $M_{0;2}^{(r)}$ and $M_{1;1}^{(r)}$ respectively (for $n=3$). 
They are typically bell--shaped
(except for $r=1$) and perhaps obey scaling relations 
similar to those of the spectral functions examined in 
ref.~\cite{JanosMax}. The figures show how they build up
the sum of moments $M_{0;2}+\Mt_{0;2}$ and $M_{1;1}+\Mt_{1;1}$.
Using the exact ratio of the mass to the $\Lambda$--parameter
\be
\frac{M}{\Lambda_\msbar}=\frac{(8/\rme)^{1/(n-2)}}{\Gamma[1+1/(n-2)]}\,,
\end{equation}
obtained by Hasenfratz, Maggiore and Niedermayer \cite{HMN},
we also exhibit the perturbative results up to and 
including terms of order $\lambda(q^2)$. The agreement of the
summed terms and PT is impressive for $-q^2/M^2\sim 10^5$.
For values of $-q^2/M^2>\sim 10^6$ contributions from states with
$\ge 7$ particles must be taken into account. Note we have also 
included the contribution of the one particle states in the sums;
these tend to improve the agreement at lower values of $-q^2$
and fall asymptotically as $M^{(1)}_{l;N}\sim m_l\pi^4/[4\ln^2(-q^2/M^2)]$
with $m_0=1,m_1=1/2,m_2=-1/2$.

\begin{figure}[t]
\begin{center} 
%\vspace*{-1mm}
\hspace{1.0cm}   
\epsfig{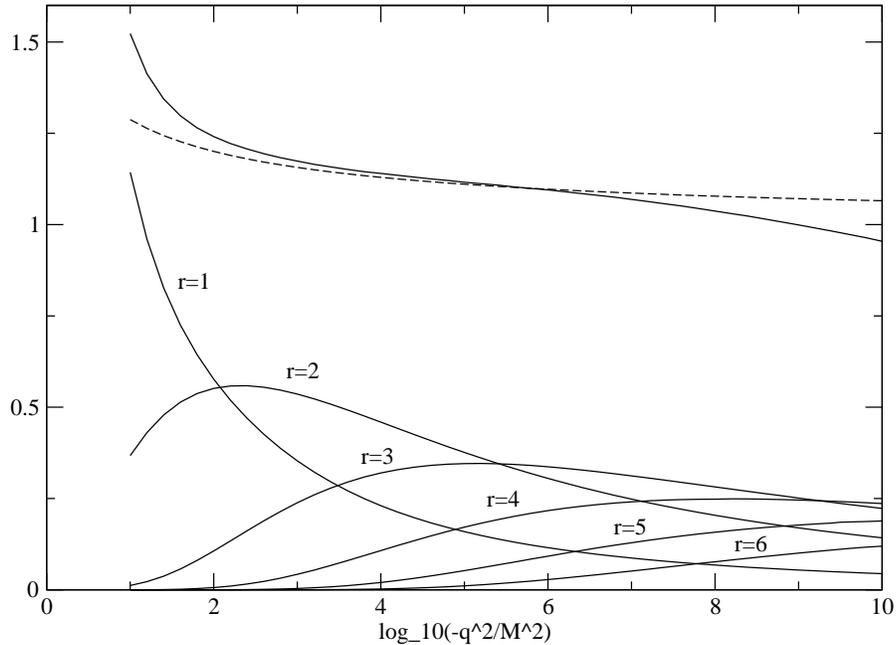}
\caption{\footnotesize
Contributions $M_{0;2}^{(r)}$ for $n=3$ from $r=1,\dots,6$--particle 
states. The upper full line is their sum.  
The dashed line is the perturbative expansion 
of $M_{0;2}+\Mt_{0;2}=1+\lambda$ 
up to and including terms of order $\lambda(q^2)$.
}
%\vspace*{-10mm}
\end{center}
\label{FigM02}
\end{figure}

\begin{figure}[t]
\begin{center} 
%\vspace*{-1mm}
\hspace{1.0cm}   
\epsfig{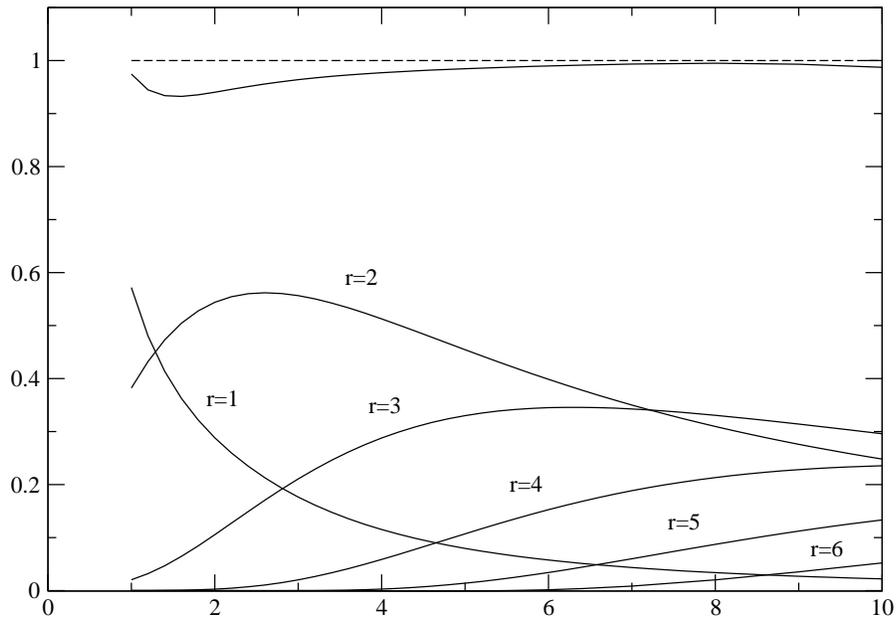}
\caption{\footnotesize
As for Fig.~5 but for the moment $l=1,N=1$; here the 
PT result is $1+\rmO(\lambda^2)$.}
%\vspace*{-10mm}
\end{center}  
\label{FigM11}
\end{figure}

\subsection{Conclusions}

Many qualitative field--theoretic features first observed in  
non--perturbative studies of integrable models in $2d$, 
have in the past found their counterparts in realistic
models in 4 dimensions. Although fascinating in their own right,
we hope that the investigations of structure functions 
of $2d$ O$(n)\,\,\sigma$--models presented in this paper will play a 
similar r\^{o}le. Similar methods are applicable to many other 
physical situations e.g. generalized structure functions, 
exclusive electro--production processes and rapidity gaps.

\subsection{Acknowledgements}

We would like to thank Ferenc Niedermayer for many instructive
discussions on the parton model, Fred Jegerlehner for sending us
data files on the QCD Adler function, Martin L\"{u}scher and
Christian Kiesling for reading the manuscript, and Vladimir Chekelian 
for information on the HERA data. 
This investigation was supported in part by the Hungarian 
National Science Fund OTKA (under T034299 and T043159).

\vfill
\eject

%%%%%%%%%%%%%%%%%%%%%%%%%%%%%%%%%%%%%%%%%%%%%%%%%%%%%

% List of references

\eject

\end{document}